\documentclass[review,number,sort&compress]{elsarticle}
\usepackage{amsmath}
\usepackage{subfig}

\begin{document}

\begin{frontmatter}

\title{Charge Measurement of Cosmic Ray Nuclei with the Plastic Scintillator Detector of DAMPE}

 \author[pmo,sass] {Tiekuang Dong \corref{cor}}
 \ead{tkdong@pmo.ac.cn}
 \author[imp] {Yapeng Zhang \corref{cor}}
 \ead {y.p.zhang@impcas.ac.cn}
 \author[pmo] {Pengxiong Ma}
 \author[imp] {Yongjie Zhang}
 \author[infn,uds] {Paolo Bernardini}
 \author[imp] {Meng Ding}
 \author[ihep] {Dongya Guo}
 \author[pmo] {Shijun Lei}
 \author[pmo] {Xiang Li}
 \author[gssi,ngs] {Ivan De Mitri}
 \author[ihep] {Wenxi Peng}
 \author[ihep] {Rui Qiao}
 \author[infn,uds] {Margherita Di Santo}
 \author[imp] {Zhiyu Sun}
 \author[infn] {Antonio Surdo}
 \author[gssi,ngs] {Zhaomin Wang}
 \author[pmo] {Jian Wu}
 \author[pmo] {Zunlei Xu}
 \author[imp] {Yuhong Yu}
 \author[pmo,sass] {Qiang Yuan}
 \author[pmo] {Chuan Yue}
 \author[pmo] {Jingjing Zang}
 \author[ustc] {Yunlong Zhang}

\cortext[cor]{Corresponding author:}

\address[pmo]{Key Laboratory of Dark Matter and Space Astronomy, Purple Mountain Observatory, Chinese Academy of Sciences, Nanjing 210008, China}

\address[sass] {School of Astronomy and Space Science, University of Science and Technology of China, Hefei,
Anhui 230026, China}

\address[imp] {Institute of Modern Physics, Chinese Academy of Sciences, 509 Nanchang Road, Lanzhou 730000, China}

\address[infn] {Istituto Nazionale di Fisica Nucleare (INFN) - Sezione di Lecce, I-73100  Lecce, Italy}

\address[uds] {Dipartimento di Matematica e Fisica E. De Giorgi, Universit\`a del Salento, I-73100 Lecce, Italy}

\address[ihep] {Institute of High Energy Physics, Chinese Academy of Sciences, Beijing 100049, China}

\address[gssi]  {Gran Sasso Science Institute, GSSI, Via M. Iacobucci 2, L'Aquila, Italy}

\address[ngs]  {INFN Laboratori Nazionali del Gran Sasso, Assergi (L'Aquila), Italy}

\address[ustc] {State Key Laboratory of Particle Detection and Electronics, University of Science and Technology of China, Hefei 230026, China}

\begin{abstract}
One of the main purposes of the DArk Matter Particle Explorer (DAMPE) is to measure the cosmic ray nuclei up to several tens of TeV or beyond, whose origin and  propagation remains a hot topic in astrophysics.  The Plastic Scintillator Detector (PSD) on top of DAMPE is designed to measure the charges of cosmic ray nuclei from H to Fe and serves as a veto detector for discriminating gamma-rays from charged particles.  We propose in this paper a charge reconstruction procedure to optimize the PSD performance in charge measurement. Essentials of our approach, including track finding, alignment of PSD, light attenuation correction, quenching and equalization correction are described detailedly in this paper after a brief description of the structure and operational principle of the PSD.  Our results show that the PSD works very well and almost all the elements in cosmic rays from H to Fe are clearly identified in the charge spectrum.
\end{abstract}


\end{frontmatter}

\section{Introduction}           %% first-level sections will be auto-capitalized
Cosmic rays are high energy nuclei that diffuse across the interstellar space. These high energy particles are supposed to be accelerated in some environments, such as supernova remnants, pulsars, and/or quasars. Although cosmic rays were discovered more than one hundred years ago, some fundamental questions regarding them remain controversial, such as the acceleration mechanism, the material components of source region, and the transportation process etc. In recent years, with the rapid development of detection techniques several balloon-borne and space-based missions have been  carried out, such as ATIC\,\cite{Guzik2004}, CREAM\,\cite{Seo2004}, PAMELA\,\cite{Picozza2007}, Fermi LAT\,\cite{Atwood2009}, and AMS-02\,\cite{Kounine2012}. Some results provided by these detectors  disagree with the theoretical predictions of current models. For example, the very recent results of AMS-02\,\cite{Aguilar2015a,Aguilar2015b}, and CREAM-III\,\cite{Yoon2017} showed that neither the flux of proton nor Helium can be described by a single power law spectrum and that the shapes of proton and Helium fluxes are different. Actually, the fluxes of proton and Helium measured by PAMELA\,\cite{Adriani2011} have clearly shown the evidence of hardening in the hundreds of GV region earlier. Previously, the CREAM\,\cite{Ahn2010} experiment reached indirectly the hardening of proton and Helium fluxes. In addition, the results from ATIC-2\,\cite{Wefel2008} also implied the different energy spectra for proton and Helium. As a space-borne particle detector, DArk Matter Particle Explorer (DAMPE) is also designed to measure cosmic ray flux with high accuracy and wide energy range\,\cite{Chang2014,Ambrosi2017}.

The detailed  structure of DAMPE can be found in the DAMPE mission paper\,\cite{Chang2017}. Here we  merely give a brief introduction of DAMPE for the convenience of following discussions. As shown in Fig.1, DAMPE consists of four sub-detectors: the Plastic Scintillator Detector (PSD), the Silicon-Tungsten tracKer-converter (STK), the BGO imaging calorimeter (BGO), and the NeUtron Detector (NUD). The PSD is a charge detector to measure the charge of nuclei as well as an anti-coincidence detector to distinguish gamma-rays from charged particles\,\cite{Xu2018}. The role of STK is to convert gamma rays into electron/positron pairs, to reconstruct the track and to measure the charge of cosmic ray nuclei. The BGO calorimeter has two main roles: one is to measure the profile of energy deposition from which one can  distinguish  electromagnetic and hadronic showers, and the other one is to measure the total energy deposition from which the primary energy can be estimated. The NUD can monitor the secondary neutrons mainly created by the hadronic shower. So NUD can be used for a redundant separation of hadronic and electromagnetic showers. Synchronized by a global trigger system, all the four sub-detectors work collaboratively, making DAMPE a powerful multifunctional high energy particle space telescope.

\vspace{-5mm}
 \begin{figure}
   \centering
   \includegraphics[width=12cm, angle=0]{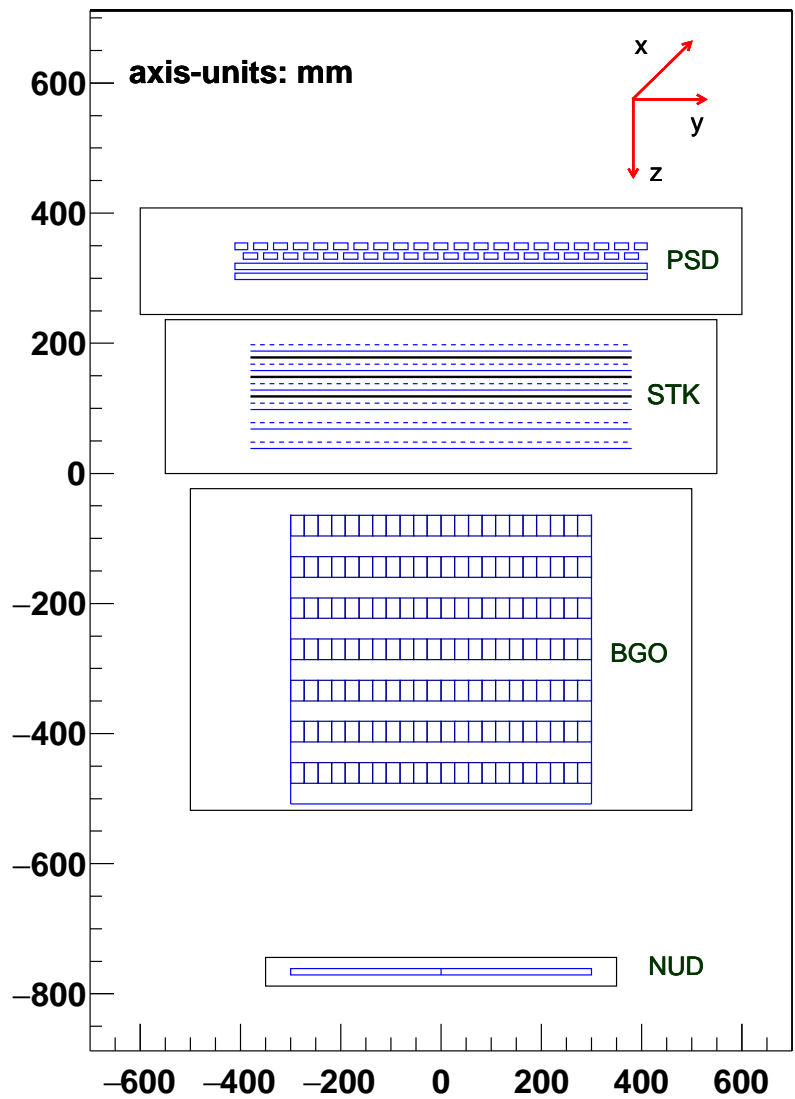}
   \caption{Diagram of DAMPE and the satellite. In order to see clearly the structure of sensitive detector modules, the satellite platform, support structure, front end electronics and so on are not shown. The detailed structure can be found in the mission paper\,\cite{Chang2017}.}
   \label{Fig1}
   \end{figure}
\vspace{5mm}

Charge measurement is the foundation of cosmic ray flux measurement.  A high charge resolution is  crucial to obtain cosmic ray energy spectra with high accuracy.
In principle, both PSD and STK can measure the charges of cosmic ray nuclei   as they are all thin detectors. However, the heavy saturation effect for STK makes it difficult to measure nuclei heavier than Oxygen ($Z=8$). Meanwhile, PSD is designed with a quite large dynamical range of energy measurement to detect the nuclei up to at least Fe \,\cite{Yu2017,Zhang2017,Zhou2016}. In this paper, we will focus on charge measurement by PSD. STK charge detection will also be considered when we analyze nuclei lighter than Oxygen. The detailed data analysis method about charge measurement by STK will be published elsewhere. In this paper we will focus on the data analysis method  about charge measurement by PSD.

\section{The Plastic Scintillator Detector (PSD) of DAMPE}
The detailed structure of the Plastic Scintillator Detector (PSD) of DAMPE can be found in Ref.\,\cite{Yu2017}. For completeness of this paper, a brief description is given. PSD is composed of two planes: the upper/lower plane is arranged along the X/Y axis in the satellite coordinate system (see Fig.1). In each plane 41 PSD bars are placed in parallel with a two-layer configuration. The dimension of a PSD bar is 884 mm long $\times$ 28 mm width (25 mm for bars at edges) $\times$ 10 mm thickness. The bars of the two layers in the same plane are staggered by 8 mm to ensure a full coverage of the detector by the active area of scintillators. With this crisscross structure, an active area of 825 mm$\times$ 825 mm is achieved and an effective incident particle would penetrate at least two bars. This arrangement can improve the detection efficiency and accuracy of charge measurement. Each end of a PSD bar is viewed by a photomultiplier tube (PMT). When a charged particle much heavier than electron passes through a PSD bar, it loses energy mainly by ionization process. On average, the energy deposition in a PSD bar is  proportional to the product of the square of its charge ($Z^{2}$) and the real path length.

\section{Charge Reconstruction Procedure}
 According to the structure of PSD, the basic idea of charge measurement by PSD is  the following: The PSD is seen as a detector with 4 layers. Each PSD bar provides two independent measurements of a single particle when a particle pass through it. But if the particle pass through the corner of a PSD bar the path length will be very small (much less than the thickness of 10 mm), and hence the relative error of the energy deposition will be large. So we only consider the charge values measured by a PSD bar when the path length is quite large. So one have at least 4 and at most 8 measurements for an incident particle. Finally, the arithmetic mean value is calculated as the best estimation of the charge.

 In order to get the charge   measurement  of a single PMT, we must determine the impact position and path length accurately. It is because the scintillating light excited by ionization energy loss in a PSD bar  attenuates before  reaching a PMT, and the attenuation proportion depends on the distance between the impact point and the PMT. So the measured energy (proportional to the light collected by a PMT) for each end of PMT also depends on the impact position. The approach of converting the recorded energy by each PMT into the real energy deposition per unit path length is called light attenuation correction. If the geometry of the PSD bars are known we can calculate the path length in every bar intercrossed by the real track. However, the real geometrical parameters of the PSD bars slightly deviate from the designed values due to several factors, such as limited installation precision, vibration during launching, weightlessness in space. Here weightlessness in space means that the gravity force is compensated by the fictitious centrifugal force on orbit. The approach of determining the real position and orientation of PSD bars is called alignment correction. Light attenuation correction and alignment correction can be performed using proton MIP events.

 Besides these two corrections, two other corrections should also be considered: the equalization correction and quenching correction. The final result of charge reconstruction is the charge spectrum, from which one can identify all the abundant elements in cosmic rays (from H to Fe). The charge spectrum is obtained by collecting the charge values of a large number of particles.  However, charge measurements from different PMTs (164 PMTs in total) may be not consistent with each other due to slightly different responses of the PMTs. If the charge values of a certain element measured by different PMTs with slightly different peak positions are collected together the charge distribution will be broadened. Therefore, we should make the peak positions of the charge spectrum measured by every PMT overlap with each other. This step is called equalization correction. The quenching correction arrives from the quenching effect, i.e., with the increasing of energy deposition density along the track of charged particles the output scintillating lights is no longer proportional to the energy deposition due to saturation  of scintillator molecules. The result is that in the uncorrected charge spectrum the peak position of every element in the charge spectrum is lower than its charge number. And the quenching effect becomes more  severe with increasing charge number. For example, the peak position of C ($Z=6$) is about 5.9, while that of Fe ($Z=26$) is only about 19.6. The  measured charge values could be corrected to the real ones by applying a quenching function which is an analytical expression described by some parameters (see below). The step of determining the parameters of quenching function is called quenching correction. Actually, the equalization correction and quenching correction are made at the same time (see below).

According to above charge reconstruction procedure, we will describe the steps in following subsections in more detail.

\subsection{Track Finding}
The first step of charge reconstruction is to find the real track of the primary particle. According to the structure of DAMPE, there are two different ways to find the track of an incident particle. One is based on the clusters induced by charged particles in STK (called STK track), and the other one is based on the shower profile in BGO calorimeter (called BGO track). The cluster in STK represents the clusters of continuous fired read-out strips. The center of the cluster ($x_{c}=\sum x_{i} ADC_{i}/\sum ADC_{i}$) provides a measurement of the impact position. The impact positions in different STK layers could be used to reconstruct the STK track. On the other hand, the shower profile represents the distribution of energy deposition in BGO bars. The main axis of the shower provides an estimation of the track in the frame work of detector.

Actually, there are usually several STK tracks for primary and secondary particles created when an incident nucleus interacts with the detector material. For example, the spallation reactions between the incident heavy nuclei and the nuclei in PSD and upper layers of STK will produce some fragments, each of which may leave an STK track. In addition, a part of secondary particles created during the hadronic shower development will move backward relative to the direction of the primary particle. According to the theory of hadronic shower, many of secondary particles are charged pions ($\pi^{\pm}$). The clusters induced by $\pi^{\pm}$ in STK are very similar to proton induced ones since they have the same absolute charge number. Heavy nuclei can be identified by looking at the ADC values of STK clusters along the track. However, it is very difficult to distinguish the primary proton and secondary $\pi^{\pm}$ just according to ADC values of STK clusters.
So we divide all the events into two classes: heavy nuclei candidates and light nuclei ones. If both the maximum energy deposition among the 41 PSD bars in the upper plane and lower plane are larger than 20 MeV the events are classified as heavy nuclei candidates. Otherwise the events are classified as light nuclei candidates. For a given heavy nucleus candidate we choose the STK track with the largest mean ADC value as the primary track. For a light nucleus candidate the tracks created by the primary particle and secondary ones are difficult to distinguish only according to the mean ADC value. In order to increase the accuracy of track finding for light nucleus candidates, the information of BGO track should be considered since usually a single BGO track (describing the shower axis) can be reconstructed for an event. Due to the low granularity of BGO bars (25 mm $\times$ 25 mm), the accuracy of BGO track can not meet the requirement of accurate charge reconstruction. Our method is to constrain the region of STK track finding around the BGO track when it is available. The STK tracks determined by this way are called global tracks. In the STK track reconstruction stage only the center of cluster ($x_{c}=\sum x_{i} ADC_{i}/\sum ADC_{i}$) is considered. So some global tracks may include the clusters induced by different particles (primary and secondaries). In order to select the real track and maintain a high selection efficiency we set a loose homogenization condition: the maximum ADC is less than 5 times of the minimum one on the global track.

\subsection{Alignment Correction}
Once the track is determined, we calculate the impact coordinates and the path length in the PSD bars of an incident particle. In calculations we need the real geometry of the PSD bars. As mentioned above, however, the real geometry of PSD bars may deviate from the designed parameters due to various factors. The detailed alignment method will be published elsewhere\,\cite{Ma2018}. Here we only give a brief introduction. It should be noted that in the alignment correction the bending of PSD bars are ignored for two reasons: (i) in order to avoid the bend of PSD bar due to temperature change, we only fix one end of each PSD bar allowing expansion or contraction on the other end\,\cite{Yu2017}; (ii) the satellite on-orbit is in the weightless state due to the compensation between the gravity force and the fictitious centrifugal force. That is to say the gravity does not deform the PSD bars.

The alignment correction is performed using proton MIP events. It is because about 90\% of cosmic rays are protons and the tracks of proton MIP events can be reconstructed very accurately. Here the proton MIP events means that the protons penetrate the BGO calorimeter without suffering any hadronic interactions with the detector material.

In order to examine the effects of misalignment on charge spectrum, we divide the length of each PSD bar uniformly into 11 segments. At first, we select some ideal events that pass through the whole PSD bar from the upper surface to the lower surface by setting a restrict geometry condition. The underlying hypothesis is that the misalignment is small. Results show that the charge distribution of selected events agree well with the one obtained by Geant4 simulation (http://cern.ch/geant4) where the geometry model without misalignment is used. It means that the estimated path length from the upper surface to the lower surface is quite accurate, implying that the rotations in $XZ$ and $YZ$ planes are negligible ($X$, $Y$, $Z$ axes are shown in Fig.1). In addition, in the event reconstruction procedure the particles that pass through the PSD bars at edges are rejected. So the small longitudinal shift in the X (Y) direction of PSD bars arranged along X (Y) axis can also be ignored. Then, there are only 3 degrees of freedom left for each PSD bar: the rotation in $XY$ plane, translation along the $Z$ axis, and transverse translation in $XY$ plane.

For each PSD bar we set 3 alignment parameters corresponding to above three degrees of freedom. For each proton MIP event that pass through the corner (which is called corner-passing event) of PSD, the path length ($PL$) can be calculated as the function about these parameters. For a given set of alignment parameters, the path length can be calculated. According to Geant4 simulation we set the most probable energy deposition per millimeter $S$ = 0.2 MeV/mm. Then we can calculate the energy deposition and make it equal to the real energy deposition:
\begin{equation}
S \times PL = E_{dep}.
\end{equation}
For $N$ corner-passing events we can get a matrix equation. This equation can be solved by iteratively looking for the least square solution until the convergent results are reached\,\cite{Ma2018}.

The final results of alignment parameters will be used for subsequent charge reconstruction. Actually, we have developed some functions in the data process software (DmpSW) using the alignment parameters to calculate the path length when the track and PSD bar is given.

\subsection{Light Attenuation Correction}
The sensitive material of the PSD is one type of organic scintillator EJ200 with the column density (density $\times$ thickness) of 1.032 g/cm$^{2}$.  The thickness of each PSD bar is only 1 cm. Due to the low mean $Z$ and thickness, PSD  can be seen as a thin detector. So the main energy transfer process is ionization energy loss when heavy charged particles pass through PSD bars. In general there is a shift between the emission spectrum and the absorption spectrum for the plastic scintillator, which induce the self-absorption (attenuation) of the scintillator light. The attenuation extent is proportional to the transmission distance. This can explain why plastic scintillators are transparent to the fluorescence photons emitted by themselves and why it is necessary to apply light-attenuation correction for long PSD bars (with the length of about 80 cm).

In the light attenuation correction, it is reasonable to assume that the rate of absorption does not depend on the light intensity. Under this hypothesis we can get the correction functions for all the PSD bars using proton MIP events and apply them to all the cosmic ray nuclei from H to Fe. By using the track of proton MIP events we can obtain the impact point and path length with the help of tools developed based on the alignment correction. We artificially divide each PSD bar into 80 segments, the length of each segment is 1 cm. If the impact point is in a given segment we calculated the charge value measured by each end of PMT and fill it into a histogram. For each PMT we get 80 histograms. Each histogram can be fitted by the Gaussian convoluted Landau distribution function.  Therefore we get the most probable value (MPV) for each segment. In other words, we get the MPV value of the charge distribution as a function of the distance between the impact point and the PMT. Since the fluorescence lights are emitted isotropically by exited molecules, the transport path of photons may be quite complex. Anyway on the average the transmission length depends on the distance between the impact point and PMT. If the light travels parallel to the PSD bar the light intensity will decay exponentially with the transmission length. Actually, the results show that, the decay of MPV value with the distance can be expressed approximately by an exponential function. However, at the largest distance the function deviates from the exponential function. Let's give a typical example. Fig. 2 shows the MPV value of charge distribution for one PMT of one PSD bar versus the distance between the impact point and the PMT, where the red line is the exponential decay function.

There are at least two methods to correct the attenuation effect. One is to interpolate the MPV value as the function of distance, the other one is to describe the attenuation behavior by a simple analytical formula. The latter is more convenient and is used in this paper. The attenuation formula used in this paper is the sum of an exponential decay function and a cubic polynomial function:

\begin{equation}
    Q_{MPV}(d) =  C_{0}e^{-d/\lambda}+C_{1}d+C_{2}d^2+C_{3}d^3,
\end{equation}
where $d$ is the distance between the impact point and the PMT, $Q_{MPV}$ is the MPV value of charge distribution for proton MIP events.

\vspace{-5mm}
 \begin{figure}
   \centering
   \includegraphics[width=12cm, angle=0]{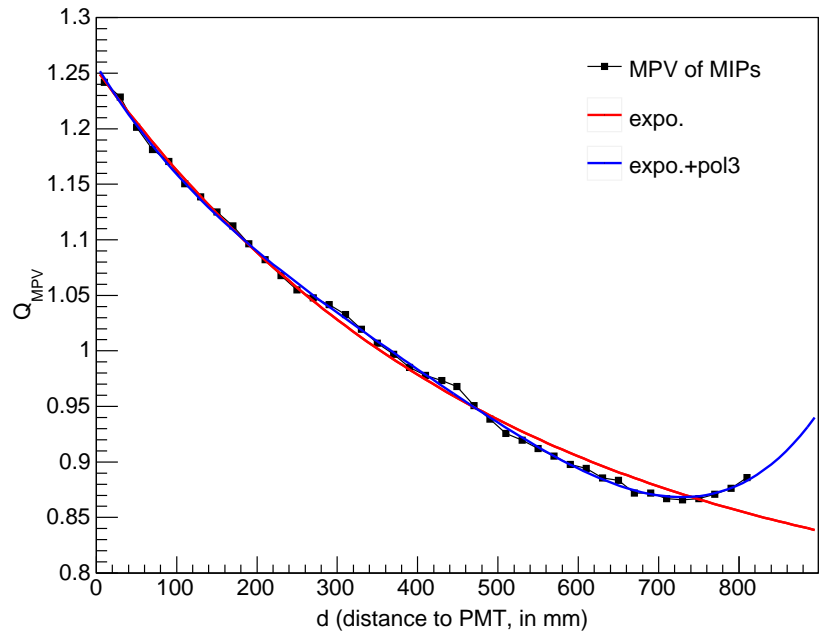}
   \caption{The MPV value of charge distribution for proton MIP events versus the distance between the impact point and the PMT. The red curve is the fit with an exponential function and the blue curve is that with Eq.(2).}
   \label{Fig2}
   \end{figure}
\vspace{5mm}

After the light attenuation correction, one  may expect that the charge values measured by two PMTs of every PSD bar will be consistent with each other since they represent the charge of the same particle. However, if the selected track is not the primary track and the impact point is wrong, the light attenuation correction will make the charge measurements of the two PMTs to be significantly different from each other. In addition, it is possible that the PSD bar is hit by more than one particle and the energy deposition is thus the sum of the energy loss of all the passing particles. Since the light attenuation correction is based on only one impact point, the charge values measured by the two PMTs also differ from each other largely. Therefore, if the charge values after light attenuation correction by PMTs reading the two ends of the same PSD bar are not consistent, the event will be rejected. This condition can improve the charge resolution further.

%Therefore, the light attenuation correction also provides a further check of the selected track. After light attenuation correction, if the charge values by PMTs reading the two ends of the same PSD bar are not consistent, the event will be rejected.  the charge resolution is further improved.
%By rejecting the events with wrongly selected track the charge resolution is further improved.

\subsection{Equalization and Quenching Correction}
Plastic scintillator can convert a fraction of the energy lost by the passed charged particles into fluorescent light. The value of this fraction is called scintillation efficiency. It is well known that the scintillation efficiency depends on the type and energy of the charged particle. The response of plastic scintillator to charged particles can be described by the relation between the energy of fluorescent light emitted per unit path length ($dL/dx$) and the specific energy loss, i.e., the energy deposition per unit path length ($dE/dx$). If the light yield is proportional to the energy loss one can get a linear response function:
\begin{equation}
    \frac{dL}{dx} = S \frac{dE}{dx},
\end{equation}
where $S$ is a normalization factor representing the scintillation efficiency.

DAMPE a high energy particle detector. We are interested in the particles above several GeV/n. In this energy region
%According to the Bethe-Bloch formula, when the energy per nucleon of a cosmic ray nucleus is larger than several GeV
the ionization energy loss is nearly a constant and proportional to the square of the charge number. It means that the ionization density along the track increases sharply with the charge number. At very high ionization density, the molecules of plastic scintillator will reduce the scintillation efficiency. This phenomenon is the so-called quenching effect.

The widely used empirical quenching formula is first suggested by Birks\,\cite{Birks1951}:
\begin{equation}
    \frac{dL}{dx} = \frac{S \frac{dE}{dx}}{1+kB\frac{dE}{dx}},
\end{equation}
where $kB$ is the quenching parameter that can be obtained by fitting experimental data. If we assume that the particle pass through a PSD bar vertically, we can get the following formula:
\begin{equation}
    \frac{L}{d} = \frac{S E/d}{1+ kB E/d }.
\end{equation}
In this equation $d$ = 1 cm is the thickness of PSD bar. So in the following equations we will omit it:
\begin{equation}
    L = \frac{S E}{1+ kB E }.
\end{equation}

As mentioned above, for high energy cosmic rays the average energy deposition per unit path length is proportional to the square of the charge number: $E = a Z^{2}$. By Geant4 simulation, we get $a$ = 2 MeV/cm for PSD bars. Then we get
\begin{equation}
    L = \frac{aS Z^{2}}{1+a kB Z^{2}}.
\end{equation}
In event reconstruction, the light yield $L$ is converted into the apparent energy deposition ($E^{*}$) by a proportional relation between them:
\begin{equation}
    L = b E^{*} = a b Z^{* 2},
\end{equation}
where $Z^{*}$ is the apparent charge number, and the coefficient $b$ can be obtained by fitting the correlation function between ADC value and the energy deposition of proton MIP events. As mentioned above, the quenching effect becomes more and more serious with the increase of charge number. So the apparent charge number $Z^{*}$ will be smaller than the real charge number for heavy nuclei. By inserting Eq.(8) into Eq.(7) we can get the following formula:
\begin{equation}
 Z^{2} = \frac{Z^{* 2}}{S/b-a kB Z^{* 2}}.
\end{equation}
However, this formula does not reproduce properly the quenching curve for the DAMPE PSD bars. Like other authors\,\cite{Craun1970} did, we modify the Birks' semi-empirical formula in order to reproduce DAMPE data better. In particular, a linear term is added into the denominator of the right side of Eq.(9):
\begin{equation}
 Z^{2} = \frac{Z^{* 2}}{S/b + c Z^{*} -a kB Z^{* 2}} = \frac{Z^{* 2}}{p_{0} + p_{1} Z^{*} + p_{2} Z^{* 2}}.
\end{equation}
It is found that this formula can describe the quenching effect very well.

\vspace{-5mm}
 \begin{figure}
   \centering
   \includegraphics[width=12cm, angle=0]{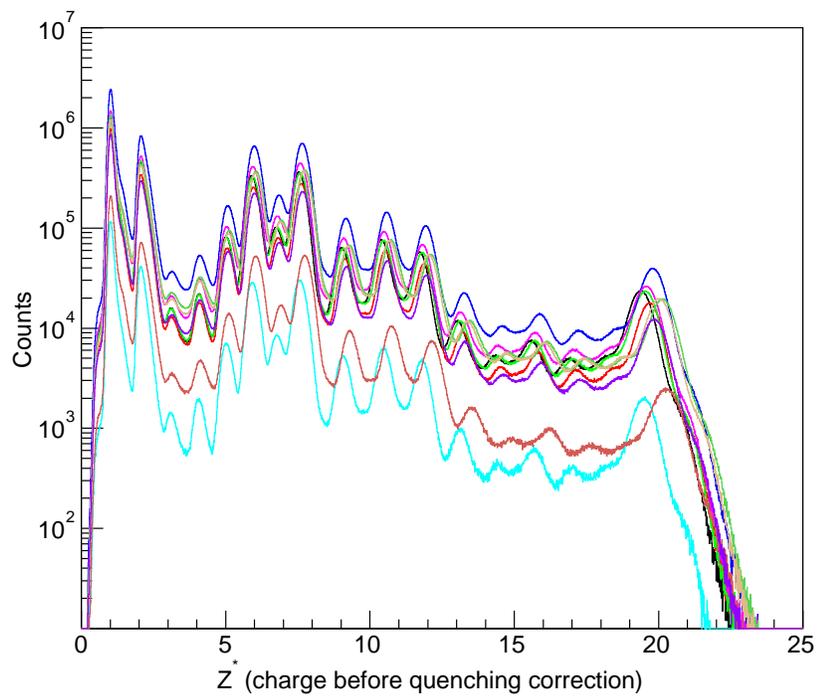}
   \caption{10 groups of charge spectra by 164 PMTs before quenching and equalization correction. For each group, the peak values of high abundant nuclei are used to fit the quenching function with Eq.(10).}
   \label{Fig3}
   \end{figure}
\vspace{5mm}

Besides the quenching effect, the equalization correction should also be considered. Although all the PSD bars are designed and produced  under the same standard, they are not identical according to the required  accuracy for the DAMPE charge measurement. In fact, we find that the peaks of charge spectra obtained by different PMTs are not always  overlapping.  Rather, we find that the charge spectra obtained by all the 164 PMTs can be roughly divided into 10 groups. Here it should be noted that the equalization and quenching correction is treated after the light attenuation correction. For each group the peaks of charge spectra are overlapping. The charge spectra for the 10 groups of PMTs are shown in Fig.3. Since the number of MPTs in these 10 groups are not equal to each other, the counts of the spectra are different. By fitting the peaks of charge for each group using Eq.(10) we get the parameters $p_{0}$, $p_{1}$, and $p_{2}$.

In principle, we can do the equalization correction of each group according to a standard, such as the spectrum of any  group. But it is more convenient to do equalization correction and quenching correction at the same time. The real charge numbers of elements just provide a natural standard.

\section{Charge Spectrum}
After making the above corrections we get the charge spectrum obtained by the whole detector. In this paper we report the results of charge spectrum based on two years of on-orbit data (from 2016-01-01 to 2017-12-31). The results are shown in Fig.4.  From this figure we can see some characters: (1) The peaks of most elements from H to Ni can be identified clearly; (2) There is the evident odd-even effect, i.e., the abundances of elements with even number of protons (even-$Z$ elements) are higher than the abundances of their odd-$Z$ neighbors. For instance, the abundances of C and O are higher than that of N, the abundances of Ne and Mg are higher than that of Na. This effect arises from the well-known pairing correlation of nucleons in nuclear physics. Some minor odd-$Z$ nuclei, such as P ($Z=15$), is partly hidden by their even-$Z$ neighbors. Even so, some odd-$Z$ elements such as P, Sc and Mn are visible; (3) The peaks for H and He are not symmetric about the corresponding peak values while the peaks for heavier nuclei are nearly symmetric about their peak values.

\vspace{-5mm}
 \begin{figure}
   \centering
   \includegraphics[width=12cm, angle=0]{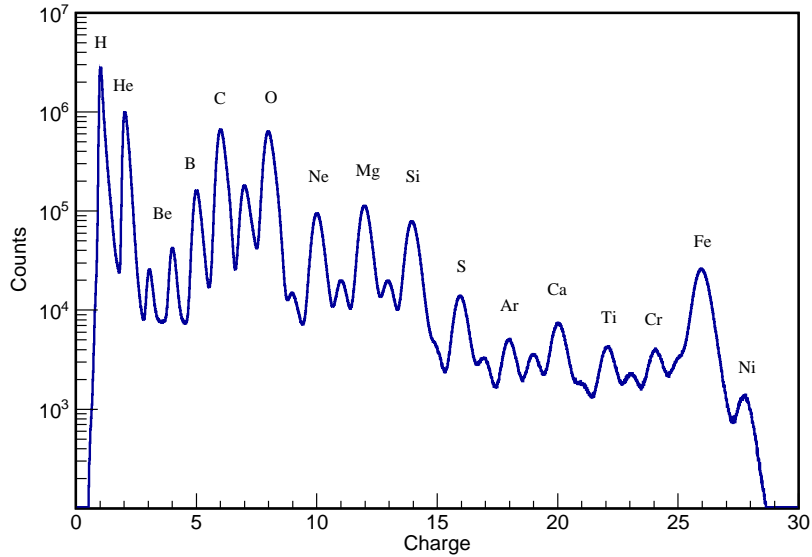}
   \caption{The charge spectrum obtained after making the alignment, light attenuation, equalization and quenching corrections. Two years of on-orbit data are used (from 2016-01-01 to 2017-12-31).}
   \label{Fig4}
   \end{figure}
\vspace{5mm}

According to the principle of interactions between heavy charged particles and thin detectors, the charge distributions of H and He can be described by the Landau distribution functions, the charge distributions of heavy nuclei can be described by the Gaussian functions. The resolutions of all the elements can be obtained by a multi-peak fitting where the peaks of proton and helium are described by the Gaussian convolved with Landau distribution functions and the peaks for heavier nuclei are described by the Gaussian functions. The odd-even effect will influence the accuracy of resolutions for low abundant odd-$Z$ elements. Therefore, we only give the resolutions for some abundant elements (see Table 1). For H and He, the values of full width at half maximum (FWHM = 2.355 $\sigma$ for Gaussian distributions) are 0.137 and 0.238, respectively.

\section{Summary and Outlook}
In this paper we give a detailed data analysis procedure  about charge measurement by PSD of DAMPE. The basic idea is that each PSD bar may provide two measurements by the two PMTs at the two ends of the bar and that the PSD is seen as a four-layer detector. Therefore, the charge of a single particle can be detected at least 4 times and at most 8 times, the mean value is used as the estimation of the charge. The main steps of charge reconstruction include the track finding, alignment correction, light attenuation correction, quenching and equalization correction. After these corrections we can obtain the charge spectrum. It is shown that almost all the elements in cosmic rays from H to Fe can be identified clearly. For example, the charge resolutions of C and Fe are 0.18 and 0.30, respectively. Such excellent charge resolution will be very useful for the elemental analysis of cosmic ray flux. Even so, more accurate corrections and more advanced methods are under considerations. For example, the light leakage at some points of PSD bars will induce the fine structure in the light attenuation curve, and the specific energy loss depends on the primary energy of particles. One can expect that the charge resolution will be improved with the improvement of analysis method.

\vspace{-5mm}
\begin{table}
%\begin{minipage}[]{200mm}
\caption[]{Charge resolution of some abundant elements.}
%\end{minipage}
\setlength{\tabcolsep}{2pt}
 \begin{tabular}{ccp{10mm}ccp{10mm}ccp{10mm}cc}
  \hline
  Element& $\sigma_Z$ && Element& $\sigma_Z$ && Element & $\sigma_Z$ && Element & $\sigma_Z$ \\
  \hline
   Li& 0.14 & &  C & 0.18 && Ne & 0.21 &&  S  & 0.25 \\
   Be& 0.21 & &  N & 0.21 && Mg & 0.22 &&  Ca & 0.29 \\
   B & 0.17 & &  O & 0.20 && Si & 0.25 &&  Fe & 0.30 \\
  \hline
\end{tabular}
\end{table}
\vspace{5mm}

\section*{Acknowledgments}
  This work is supported by National Key Research and Development Program of China (No. 2016YFA0400200), and by the National Natural Science Foundation of China (Grant Nos. U1738205, 11673075, 11643011, 11673047, U1631111, and 11761161001), and by the Strategic Priority Research Program of the Chinese Academy of Sciences (No. XDB23040000). In Europe the work is supported by the Italian National Institute for Nuclear Physics, and the Italian University and Research Ministry.

%\section*{References}


\begin{thebibliography}{99}
%% you can type \apj for ApJ, \aap for A&A, \apss for Ap&SS, etc. Please consult
%% the macro chjaa.cls. You can also find them in aasguide.tex (AASTeX for ApJ, AJ, PASP)
%% Please follow the format of ChJAA's reference list
\bibitem{Guzik2004} T. G. Guzik, J. H. Adams, H. S. Ahn et al., Adv. Spa. Res. {\bf 33}(10) (2004) 1763.
\bibitem{Seo2004} E. S. Seo, H. S. Ahn, J. J. Beatty et al., Adv. Spa. Res. {\bf 33}(10) (2004) 1777.
\bibitem{Picozza2007} P. Picozza, A. M. Galper, G. Castellini et al., Astropart. Phys. {\bf 27} (2007) 296.
\bibitem{Atwood2009} W. B. Atwood, A. A. Abdo, M. Ackermann et al., Astrophys. J. {\bf 697} (2009) 1071.
\bibitem{Kounine2012} A. Kounine, Int. J. Mod. Phys. E {\bf 21}(8) (2012) 1230005.

\bibitem{Aguilar2015a} M. Aguilar, D. Aisa, B. Alpat et al. (AMS Collaboration), Phy. Rev. Lett. {\bf 114} (2015a) 171103.

\bibitem{Aguilar2015b} M. Aguilar, D. Aisa, B. Alpat et al. (AMS Collaboration), Phy. Rev. Lett. {\bf 115} (2015b) 211101.

\bibitem{Yoon2017} Y. S. Yoon, T. Anderson, A. Barrau et al., Astrophys. J.  {\bf 839} (2017) 5.

\bibitem{Adriani2011} O. Adriani, G. C. Barbarino, G. A. Bazilevskaya et al., Science {\bf 332} (2011) 69.

\bibitem{Ahn2010} H. S. Ahn, P. Allison, M. G. Bagliesi et al., Astrophys. J. Lett. {\bf 714} (2010) L89.

\bibitem{Wefel2008} J. P. Wefel, J. H. Adams, H. S. Ahn et al., Proceedings of the 30th Inernational Cosmic Ray Conference {\bf 2} (2008) 31.

\bibitem{Chang2014}  J. Chang, Chin. J. Space Sci. {\bf 34}(5) (2014) 550.

\bibitem{Ambrosi2017} G. Ambrosi, Q. An, R. Asfandiyarov et al. (DAMPE collaboration), Nature {\bf 552} (2017) 63.
\bibitem{Chang2017}  J. Chang, G. Ambrosi, Q. An et al., Astropart. Phys. {\bf 95} (2017) 6.
\bibitem{Xu2018} Z. L. Xu, K. K. Duan, Z. Q. Shen et al., RAA {\bf 18} (2018) 27.
\bibitem{Yu2017} Y. H. Yu, Z. Y. Sun, H. Su et al., Astropart. Phys. {\bf 94} (2017) 1.
\bibitem{Zhang2017}  Y. P. Zhang, Y. J. Zhang, T. K. Dong, P. X. Ma, Y.H. Yu, P. Bernardini(DAMPE collaboration), 35th International Cosmic Ray Conference PoS (ICRC2017) 168.
\bibitem{Zhou2016} Y. Zhou, Z. Y. Sun, Y. H. Yu et al., Nucl. Instr. Meth. Phys. Res. A {\bf 827} (2016) 79.

\bibitem{Ma2018} P. X. Ma, Y. J. Zhang, Y. P. Zhang et al., arxiv:1808.05720.

\bibitem{Birks1951} J. B. Birks, Proc. Phys. Soc. A {\bf 64} (1951) 874.

\bibitem{Craun1970} R. L. Craun and D. L. Smith, Nucl. Instrum. Meth. {\bf 80} (1970) 239.
\end{thebibliography}
\end{document}